# Protecting and Enhancing the Photoelectrocatalytic Nitrogen Reduction to Ammonia Performance of InGaN Nanowires using Mo$_2$C Nanosheets and GaN Buffer Layer

Paulraj Gnanasekar[ab#], Karthik Peramaiya[c#], Huafan Zhang[a#], Tien Khee Ng[a], Kuo-Wei Huang[c]*, Jeganathan Kulandaivel[b]* and Boon S. Ooi[a]*

[a] Photonics Laboratory, Computer, Electrical and Mathematical Sciences and Engineering Division, King Abdullah University of Science and Technology (KAUST), Thuwal 23955-6900, Saudi Arabia

[b] Centre for Nanoscience and Nanotechnology, Department of Physics, Bharathidasan University, Tiruchirappalli-620 024, Tamil Nadu, India

[c] KAUST Catalysis Center, King Abdullah University of Science and Technology (KAUST), Thuwal 23955-6900, Saudi Arabia

\# Contributed Equally

\* Corresponding Authors

## Abstract

Photoelectrocatalytic (PEC) reduction of N$_2$ to ammonia (NH$_3$) is emerging as the potential alternative to overcome the standard Haber-Bosch approach. In this communication, solar N$_2$ reduction was demonstrated with molybdenum carbide (Mo$_2$C) co-catalyst assisted indium gallium nitride (InGaN) nanowires. The effect of aiding Mo$_2$C on InGaN NWs arrests the dark current and demonstrated the saturation current under illumination was briefly elucidated. Large NH$_3$ production of 7.2 µg·h$^{-1}$·cm$^{-2}$ with high Faradaic efficiency of 12.6 % was realized at -0.2 V vs. reversible hydrogen electrode for the Mo$_2$C/GaN/InGaN heterostructure. Notably, the proposed heterostructure also exemplifies excellent stability and reproducibility with excellent selectivity in the long-term chronoamperometry analysis. Further, the incorporation of GaN buffer layer in between Mo$_2$C and InGaN NWs was deeply investigated. From Density Functional Theory (DFT) analysis, the incorporation of GaN buffer layer aids the suitable band edge position for the transfer of photogenerated charge carrier from InGaN to Mo$_2$C co-catalyst, and unique 3d orbital of Mo$_2$C is highly suitable to hold N$_2$ for effective reduction to NH$_3$.

1. **Introduction:**

Ammonia ($NH_3$) is a vital small molecule with inflated demand in the agricultural and industrial sectors, such as manufacturing fertilizers, explosives, reagents, and resins. Besides, $NH_3$ possesses an energy density of 4.3 kW·h$^{-1}$, which also holds the possibility of the storage/transportation of hydrogen fuel.[1] Due to the indispensable needing for $NH_3$ in many industries, it is the globally second largely produced chemical and is quantified to be ~300 million tons/year. Even though the atmospheric availability of $N_2$ is abundant, the high chemical inertness of N≡N bonds (bond energy: 940.95kJ Mol$^{-1}$) leads to the kinetically challenging multiple-step $N_2$ to $NH_3$ conversion pathway.[2] As of now, the traditional Haber-Bosch approach is in action to satisfy global $NH_3$ requirements. However, the energy consumption in the Haber-Bosch process is sky-high as the process demands extremely high pressure (>150 bar) and temperature (>450 °C).[2][3] In addition, the synthesis of $H_2$ feedstock for $NH_3$ production and the Haber-Bosch process causes the massive $CO_2$ emission, which accounts for ~2% of the global emission accompanied by 1-3% of annual energy consumption. As the result of concerns in the traditional Haber-Bosch process, it is foremost essential to develop alternative energy-efficient approaches for $NH_3$ synthesis under ambient conditions. To be pinpointed, direct conversion of $N_2$ into $NH_3$ in an aqueous solution, in which water can be a proton source, is straight forward way to neglect $CO_2$ emission accompanied in the $H_2$ feedstock production and Haber-Bosch process.

Various approaches have been established including photochemical, electrochemical and photoelectrochemical (PEC) for $N_2$ reduction reaction (NRR) to $NH_3$ conversion by using renewable solar energy and water as a proton source.[4][5][6] However, PEC NRR has gained much interest among these approaches as it advances the advantages of both electrochemical and photochemical processes. Even though the PEC approach is adopted extensively for water splitting and $CO_2$ reduction reactions (Co$_2$RR), the restricted numbers of semiconductor photoelectrodes such as Si, $BiVO_4$, $WO_3$, and $TiO_2$ are reported for NRR.[7][8][9][4] Besides, the well-known III-V semiconductors are also unexplored for the PEC NRR applications. Fascinatingly, III-V semiconductors such as InGaN (Indium Gallium Nitrate) have been recognized as one of the ideal photoelectrodes for solar hydrogen evolution reaction (HER) due to their exciting properties, such as tunable bandgap over the entire visible range of the solar spectrum (0.7-3.4 eV), superior radiation resistance, excellent carrier mobility, good thermal stability, and high absorption coefficient. Moreover, the one-dimensional (1D) confined nanowire (NW) structure holds the rich solar absorption cross-section and high catalyst loading sites as compared to planar counterparts for superior PEC performance.[10][11][12,13] In addition, the 1D NW enhances the lifetime of photogenerated charge carriers that promotes rapid charge separation.

Despite the favorable properties of InGaN NWs for effective PEC, integration of efficient co-catalysts is another major challenge to achieving high solar to energy conversion of any PEC reaction. More specifically, compared to HER and $CO_2$RR, co-catalyst engineering plays a vital role in NRR as the stable N≡N bonds need to be activated on the catalyst surface for $NH_3$ production.[14] In addition, the low solubility of $N_2$ in the electrolytes increases the counterpart HER rate than NRR due to low accessible $N_2$ molecules on the electrode surface. Previously, few studies were reported on catalyst engineering to enhance the $N_2$ adsorption on the catalyst surface by creating electron-deficient sites, defects/vacancies, and preparation of porous catalyst with noble metals, transition metal chalcogenides, nitrides and carbides catalysts. Among them, molybdenum carbide ($Mo_2C$) of the transition metal carbides (TMC) family attains wide research interest owing to the peculiar electronic structure suitable for $N_2$ absorption, optimum Gibbs free energy, super stability in the electrolyte, and rich catalytic active sites. Moreover, the unoccupied *d* orbital of $Mo_2C$ is a benefit for weakening the strong N≡N and also helps to inject the electrons onto the orbital of $N_2$.[15][16][17] Notably, $Mo_2C$ nanosheets were extensively studied as NRR and HER electrocatalyst. However, the effect of $Mo_2C$ as a PEC co-catalyst is still not explored.

In this communication, we proposed 2D $Mo_2C$ protected GaN/InGaN NWs heterostructure for PEC NRR applications. The maximum Faradic efficiency of 12.6 % and maximum yield of 7.2 µg·h$^{-1}$·cm$^{-2}$ at the applied bias voltage of -0.2 V vs. RHE (Reversible Hydrogen Electrode) was realized by the proposed $Mo_2C$/GaN/InGaN NWs heterostructure. This GaN/InGaN NWs ladder structure holds the light absorption layer and a buffer layer to transfer the charge carriers towards $Mo_2C$ for effective $N_2$ reduction. Notably, the effect of GaN cladding on InGaN NWs was studied by XPS band edge position and it was found to possess the optimum band edge position to transfer the photoelectrons towards $Mo_2C$ for significant NRR performance. However, the bandgap of GaN is not in the visible region to perform the solar PEC analysis. Further, density functional theory (DFT) and first-principal calculations were used to understand the charge transfer and Gibbs free energy of the $Mo_2C/GaN/InGaN$ heterostructure for $N_2$ reduction.

**2. Methods and materials:**

**Synthesis of InGaN NWs:**

The InGaN NWs were prepared using the plasma-assisted molecular beam epitaxy (PA-MBE) approach on the medium doped n-Si (100) substrate. First, GaN seeds nucleated on the substrate under nitrogen conditions and are vertically elongated as GaN nanowire templates at an average length of 20 nm. Then, 500 nm long p-InGaN NWs were grown upon the GaN templates with an Mg-doping effusion cell temperature at 350 $^{o}C$ for 2 hrs. The p-InGaN segments behave as effective visible-light conversion mediums for NRR. Finally, an Mg-doped GaN cladding layer (~ 350 nm) was grown under Ga-rich conditions for 2 hrs to promote lateral growth.

**Synthesis of 2D $Mo_2C$ nanosheets:**

The $Mo_2C$ was purchased from Sigma Aldrich. and sonicated with DI water for 5 hours. The ultrasonication process was used to prepare the $Mo_2C$ nanosheets from bulk. 1g of $Mo_2C$ powder was added to a beaker and sonicated for 5 hours and the exfoliated $Mo_2C$ nanosheets were extracted by centrifuge at 10,000 rotations per minute (RPM) for 30 minutes. Then, the obtained exfoliated $Mo_2C$ nanosheets were dried under vacuum for 5 hours at 60 $^{o}C$.

**Removal of nitrogen contamination from $Mo_2C$ nanosheets:**

The nitrogen contaminations from the $Mo_2C$ nanosheets were removed by repetitive washing using 0.1 M HCl. Finally, the obtained $Mo_2C$ nanosheets were dried under vacuum for 24 hours and used for further investigations. Further, the absence of ammonia/nitrogen contaminations was confirmed by the indophenol blue test.

**Fabrication of $Mo_2C$ /InGaN heterostructure:**

The processed Mo$_2$C nanosheets (5 mg) were mixed with 1 ml of solution (0.3 ml of DI water, 0.68ml of ethanol, and 20 μl of Nafion) and ultrasonicated for 2 hours. Finally, the 25 μl of the prepared solution was manually dropped cast upon InGaN NWs (1 cm$^2$) and dried under vacuum at the temperature of 60 °C for 4 hours. The thickness optimization of Mo$_2$C nanosheets on the InGaN NWs was discussed in supporting information. Further, the prepared Mo$_2$C/InGaN NWs samples were provided with Ni strip contact for PEC measurements with silver paste (Pelco conductive silver paste) and Indium paste (Indium corporation, USA). Finally, to ensure the PEC device performance, we isolated the fabricated photoelectrode with epoxy (EA 9560).

**Material characterizations:**

Field emission scanning electron microscopy (FESEM) imaging was carried out with Zeiss Merlin, and Transmission electron microscopy (TEM) analysis was analyzed using FEI Titan ST. Elemental mapping analysis was witnessed with Energy dispersive X-ray spectroscopy using Oxford instruments Aztec. Photoluminescence (PL) spectroscopy was examined using the WITec Apyron PL system under two different excitation wavelengths at 325 nm and 473 nm. X-ray diffraction (XRD) patterns were acquired using a Bruker D8 Advance diffractometer equipped with a monochromator using Cu Kα (λ = 1.5405 Å) radiation. XPS studies were carried out in a Kratos Axis Ultra DLD spectrometer equipped with a monochromatic Al Ka X-ray source (hv = 1486.6 eV) operating at 75 W, a multi-channel plate, and a delay line detector under a vacuum of ~10$^{-9}$ mbar. Binding energies were referenced to the C 1s binding energy of adventitious carbon contamination which was taken to be 284.8 eV. The Mo$_2$C-InGaN sample for XPS measurements was prepared by spin-coating a 5 nm-thick 5 mg/ml Mo$_2$C in 70% ethanol on InGaN NWs. Another set of samples: as-grown p-type GaN thin film, 5 nm-thick Mo$_2$C/p-type GaN, and 50 nm Mo$_2$C thin film was investigated for the Mo2C/GaN interfacial band alignment analysis. The XPSPEAK 4.1 was used for peak fitting. All XPS scan matched with Gaussian-Lorentzian functions.

**PEC analysis:**

Entire PEC analysis was carried out using the Biologic SP-150 Potentiostat. An AM 1.5G filter assisted Xe 100 W lamp was used as the light source (Asahi Spectra). For all analysis, H-type quartz electrochemical cells with Mo$_2$C/InGaN NWs as working electrode, graphite rod as the counter electrode, and Ag/AgCl as the reference electrode were used. All PEC analyses was carried out in 0.05 M H$_2$SO$_4$ with a reversible hydrogen electrode (RHE) correction using the Nernst equation

$$E_{RHE}: E_o + E_{Ag/AgCl} + 0.059*(pH) \text{ ---(eq. 1)}$$

where $E_o$ is the measurement experimental potential and $E_{Ag/AgCl}$ is the standard potential (0.197 V) of Ag/AgCl saturated KCl reference electrode at 25 °C.[18] Entire PEC measurements were carried out under the lab temperature of 25 °C. Electrochemical Impedance Spectroscopy (EIS) analysis was carried out at the applied bias of -0.2 V vs. RHE with the range of 0.1 to 10$^4$ Hz. High pure nitrogen (99.9995%) and argon (99.9995%) were used as gas sources and the flow rates were precisely regulated with mass flow controllers and fixed to be 10 SCCM.

**NH$_3$ quantification:**

The NH$_3$ yield was quantified by the indophenol blue method. Briefly, 2 ml of the post-reaction electrolyte was mixed with 2mL of coloring solutions 1 M NaOH containing 5 wt% of C$_6$H$_5$Na$_3$O$_7$.2H$_2$O and C$_7$H$_6$O$_3$. In addition to this, 1 mL of 0.05 M NaOCl and 0.2 mL of C$_5$FeN$_6$Na$_2$O (1 wt%) were added. The final solution was slowly agitated for 30 to 40 sec and kept for 0.5 h to appear green color. After that, absorbance at 650 nm was measured for the reaction solution using UV–vis analysis. Besides, the standard calibration curve was created with the known concentration of ammonium chloride solution (Figure S3). From this standard plot, the production yield of NH$_3$ can be calculated.

**N$_2$H$_4$ quantification:**

The formation of N$_2$H$_4$ was determined using the Chrisp and Watt method. Typically, 2 ml of the N$_2$H$_4$.H$_2$O is mixed with the coloring reagent (0.4g of C$_9$H$_{11}$NO in 20 ml of ethanol and 2 ml of Conc. HCl) in various concentrations and kept undisturbed for 15 minutes, and the N$_2$H$_4$ standard estimation curve was provided in figure S4.

**Faradaic efficiency (FE) calculations:**

The standard NH$_3$ determination was carried out using NaCl$_4$, as shown in figure S3. The NH$_3$ yield and FE were calculated using the equations

$$r(NH_3) = \frac{[NH_3] \times V}{T \times A} \text{---(eq. 2)}$$

$$FE = \frac{3 \times F \times [NH_3] \times V}{M_{NH3} \times Q} \text{---(eq. 3)}$$

Where [NH$_3$] is the volume of NH$_3$ in 1 ml of electrolyte, V is the electrolyte volume, T is the CA reaction time, A is the electrode active area, M$_{NH3}$ is the molecular weight of ammonia, and Q is the total charge through the electrode.

## 3. Results and discussion:

Towards the goal, InGaN NWs were synthesized on Si substrate with GaN cladding layer using PA-MBE. The FESEM images reveal the uniform formation of vertical InGaN NWs with expanded GaN tops. The average radius and height of the slender InGaN stems are 60 and 500 nm, respectively, while that of the clad GaN tops are 120 and 350 nm. (Figure 1a and S5). Notably, this coalesced GaN was expected to aid the uniform deposition of Mo$_2$C and also enrich the charge transfer to M0o$_2$C cocatalysts. On the other hand, the uniform deposition of the Mo$_2$C layer over the GaN/InGaN NWs was witnessed with an average thickness of 6 nm (Figures 1b and c). Further, the high-resolution TEM analysis of reveals the (000$\underline{1}$) direction-oriented growth of InGaN NWs, and the fast Fourier transform (FFT) pattern re-ensures its crystalline nature (inset of Figure 1d). Most importantly, Mo$_2$C nanosheets evenly cover the GaN/InGaN NWs (Figure 1e) thanks to the GaN buffer layer with an average sheet radius of 50 nm (Figure 1f) and interplanar spacing of 0.215 nm (Figure 1g). A circular ring FFT pattern of Mo$_2$C/GaN/InGaN NWs can be attributed to the multiple orientations of Mo$_2$C nanosheets (Figure 1h). The FESEM corresponding EDS mapping (figure 1i-n and S6) reveals the uniform presence of Mo$_2$C nanosheets over the InGaN NWs array.

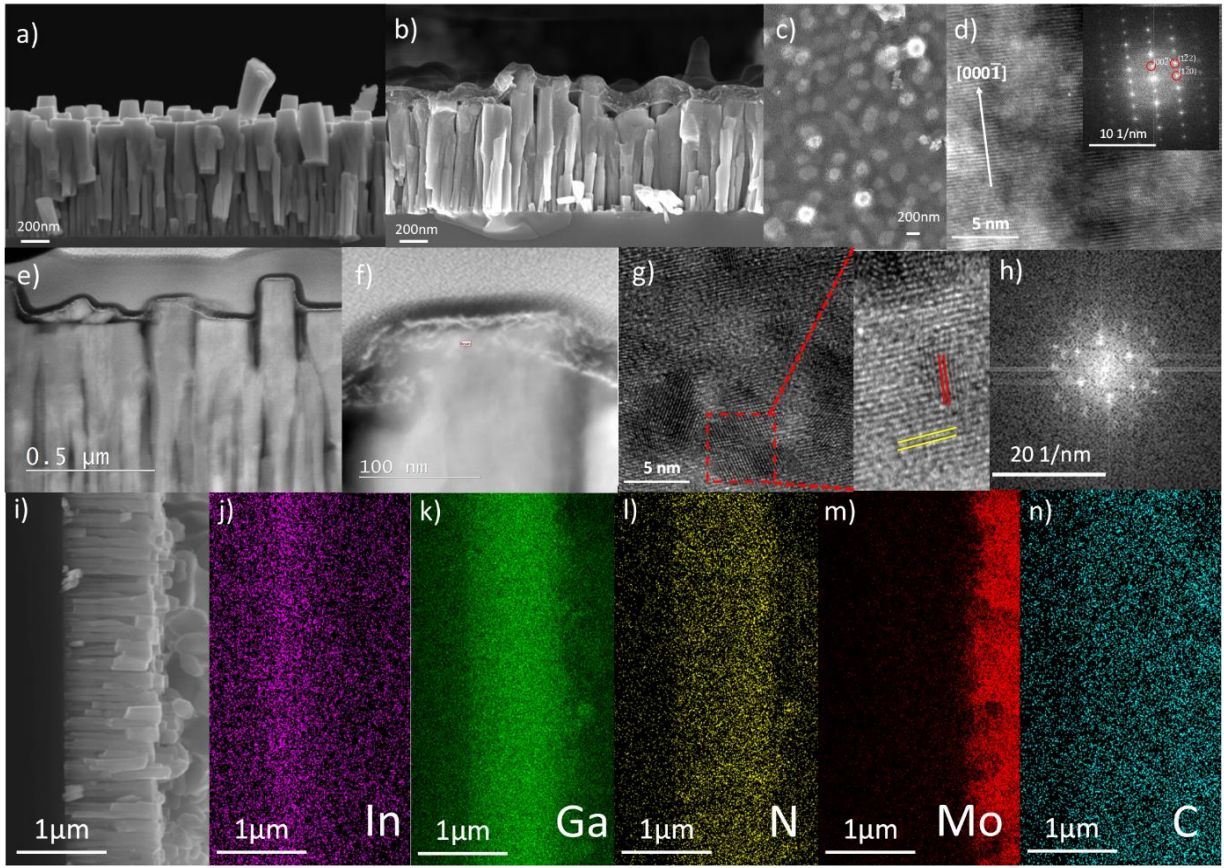

Figure 1: a and b) FESEM image of GaN/InGaN NWs and Mo$_2$C/GaN/InGaN NWs, c) Top view FESEM image of Mo$_2$C/GaN/InGaN NWs, d) HRTEM micrograph of GaN/InGaN NWs, e and f) TEM images of Mo$_2$C/GaN/InGaN NWs, g) HRTEM image of Mo$_2$C/GaN/InGaN NWs, h) FFT pattern of Mo$_2$C/GaN/InGaN NWS, i-n) EDS Mapping of Mo$_2$C/GaN/InGaN NWs.

XRD patterns of standalone GaN/InGaN NWs exhibit the characteristic 2θ diffraction peaks of GaN (002) plane at 34.5° and InGaN (100), (002), and (013) planes at 32.9°, 33.7°, and 61.7°, respectively. The patterns attribute to the InGaN stem and the coalesced top layer. After coating the Mo$_2$C, additional diffraction patterns of 2θ at 38.0°, 39.5°, and 52.2° appeared, which can be related to the characteristic Mo$_2$C planes of (002), (101) and (102), respectively, matching well with the earlier reports and JCPDS database of Mo$_2$C (35-0787).[19,20] The XRD scan confirmed the polycrystalline Mo$_2$C existence on InGaN NWs without structural modifications to the NW crystals.

The room temperature photoluminescence (PL) spectrum was obtained using two excitation wavelengths at 325 and 473 nm, as given in Figure 2b. Under the illumination of a 325 nm laser, a peak located around 363 nm (~ 3.42 eV) can be observed for both GaN/InGaN NWs and Mo$_2$C/GaN/InGaN NWs, originating from the clad GaN top. The small shoulder at ~ 390 nm may attribute to the In diffusion at the interface of the InGaN stem and clad GaN. It is worth noting that the top GaN segment absorbs the major part of the incoming light, hindering photon absorption of the InGaN stems, and thus no obvious PL emission can be observed from the InGaN stems under the excitation at 325 nm (full PL spectrum in figure S7). Therefore, PL spectrum was also recorded with a 473 nm laser source to check the bandgap of the InGaN stems. A broad peak at the visible region of the solar spectrum centred at ~620 nm and it is corresponding to the wavelength of 2 eV, which ensures an efficient solar spectrum absorption by the InGaN host. Similar PL spectra was observed for the $Mo_2C$/GaN/InGaN with a peak centred at ~ 610 nm. The blue shift of the broad band-edge emission peak may be caused by In clustering in nanowire structures or inevitable composition variation across the wafer. On the other hand, the quench in emission intensity suggests the rapid carrier separation and transfer by the $Mo_2C$ coating.

Further, XPS binding energy (BE) survey spectrum of $Mo_2C$/ InGaN indicates the presence of In, Ga, N, Mo, and C (figure S8), and XPS spectrum deconvolution results are given in Table S1.[21] Figure 2 c, d, and e present the high-resolution BE spectrum of N 1s, In 3d, and Ga 2p core levels of the as-grown GaN/InGaN NWs and $Mo_2C$/GaN/InGaN, respectively. The N 1s spectrum of $Mo_2C$/GaN/InGaN sample (figure 2c) can be decoupled into three components, corresponding to N-Ga bonds (397.0 eV), N-In bonds (395.8 eV), and Ga LMM auger line (394.0 eV), respectively. The In 3d core-level spectrum contains one doublet situated at 444.3 eV and 451.9 eV, corresponding to the In $3d_{5/2}$ and In $3d_{3/2}$ spin-orbit split components (Figure 2d), while the Ga 2p spectra are composed of Ga $2p_{3/2}$ and Ga $2p_{1/2}$, locating at 1116.9 eV and 1143.7 eV, respectively (Figure 2e). The above-mentioned peaks showed similar peak distribution as those of the as-grown GaN/InGaN NWs with negligible energy shifts, indicating the maintained InGaN crystal quality after coating. The C 1s core levels of the $Mo_2C$/ InGaN can be resolved into five peaks, which can be attributed to C-Mo bonds (284.2 eV), C-C bonds (284.8 eV), C-O bonds (286.4 eV), and O-C=O bonds (288.9 eV), as shown in figure 2e. The Mo 3d spectrum was shown in figure 2f, which can be assigned to six peaks at 228.4, 231.5, 229.3, 232.6, 233.5, 236.0 eV, related to $Mo_2C$ $3d_{5/2}$, $3d_{3/2}$, $MoO_2$ $3d_{5/2}$, $3d_{3/2}$, and $Mo_2O_3$ $3d_{5/2}$, $3d_{3/2}$, respectively. The increased C-Mo component in C 1s and the appearance of the Mo 3d peak confirm the presence of $Mo_2C$ nanosheets coating onto the InGaN NWs. [22–25] All the morphological and spectroscopical analysis confirms the formation of $Mo_2C$ nanosheets on InGaN NWs.

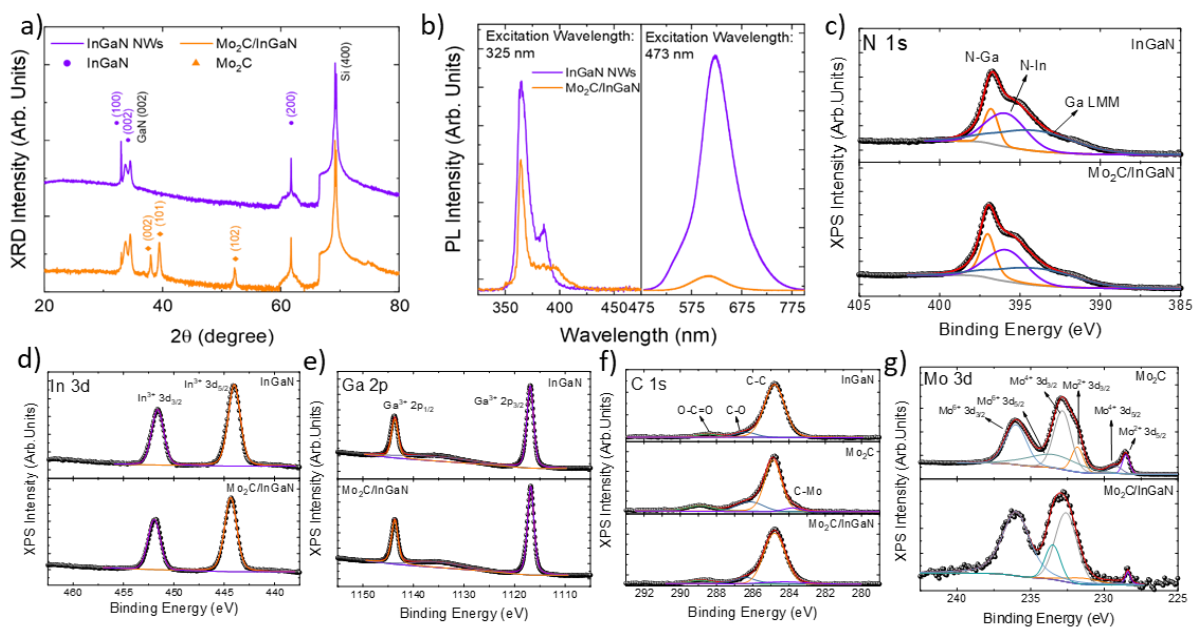

Figure 2: a) XRD pattern and b) Room temperature PL spectrum, c-g) core level binding energy spectrum of N 1s, In 3d, Ga 2p, C 1s, and Mo 3d in GaN/InGaN and Mo$_2$C/GaN/InGaN NWs heterostructure.

**PEC NRR analysis:**

Initially, the effect of the incorporation of Mo$_2$C co-catalyst in the PEC performance of GaN/InGaN NWs was evaluated by linear sweep voltammetry (LSV) analysis under the ambient condition with 0.05 M H$_2$SO$_4$ as electrolyte. Pristine GaN/InGaN NWs exhibits onset potential (potential to attain 0.1 mA/cm$^2$) of -0.23 and -0.63 V vs. RHE under illumination and dark condition, respectively (Figure 3b). In contrast, the Mo$_2$C/InGaN NWs display an onset potential of -66 mV vs. RHE and saturation current of -1.38 mA/cm$^2$ at -0.65 V vs. RHE under illumination with almost zero dark current. Furthermore, the PEC performance of Mo$_2$C/GaN/InGaN was ensured by LSV under chopped illumination, and the results are well in match with dark and light J-V graphs (Figure 3c). The Nyquist plot evidence the superior charge transfer nature of Mo$_2$C/GaN/InGaN NWs heterostructure compared to pristine InGaN NWs (Figure S9). It is clear that the Mo$_2$C nanosheets co-catalyst arrests the uncompromised dark current which can be related to the etching of InGaN NWs by harsh electrolyte and also affords the necessary active sites for superior HER performance with effective transfer of charge carriers towards the electrolyte.[26][27]

Initially, PEC nitrogen reduction performance was evaluated with LSV under Ar and $N_2$ saturated conditions, and corresponding J-V graphs are provided in Figures 3c and S10. As compared to LSV under Ar saturated condition, a measurable shift in onset potential and maximum current density is noticed under $N_2$ saturation, and this directly signposts the possibility of $N_2$ reduction at the surface of the photocathode (Figure 3d and S10).[4] Furthermore, chronoamperometry (CA) studies were performed at different applied potentials to achieve the maximum $NH_3$ production yield and Faradaic efficiency (FE) under $N_2$ saturated condition (Figure 3e). The $NH_3$ production yield in the post electrolyte solution is determined through the Indophenol-Blue method (figure S11), and the calibration data is provided in figure S3. Figure 3f and Table S2 show the $NH_3$ production yields and FE at different applied potentials. It can be seen that the maximum FE of 12.6% and $NH_3$ production of the yield of 8.9 µg/h/cm$^2$ are achieved at the applied bias of -0.2 V and -0.3 V vs. RHE, respectively. Notably, the decrease in the trend of FE over the potential of -0.2 V vs. RHE and can be related to the accelerating trend of $H_2$ production by rapid two-electron reaction rather than the six-electron. In addition, the selective $NH_3$ production by PEC NRR was confirmed by estimating hydrazine ($N_2H_4$) as it is the undesired product in the NRR. Almost zero production of $N_2H_4$ was witnessed in Watt-Chrisp colorimetric assay, indicating the absence of $N_2H_4$ formation and superior selectivity towards $NH_3$ production (figure S12). Apart from this, various control experiments were carried out to ensure that the $NH_3$ was produced by NRR, feeding by $N_2$, not from other sources. The CA test at -0.2 V vs RHE under Ar and $N_2$ saturation conditions reveal the NRR takes place by feeded $N_2$ (figure S13).

Further, the $NH_3$ production by PEC NRR was confirmed with $^{15}N_2$ isotope labeling experiment. $^{14}N_2$ gas was used as the nitrogen source for PEC NRR and the CA test was carried out at -0.2 V vs. RHE. The 1H NMR analysis of $^{14}N_2$ gas feeded electrolyte imparts the three peaks for $^{14}NH_3$ production (figure 3g). Then, to further confirm the PEC NRR, the CA test was carried out under identical conditions, except by feeding $^{15}N_2$ as the nitrogen source. As expected, the two characteristic peaks related to $^{15}NH_4^+$ were observed along with trace $^{14}NH_3$ peaks. This negligible presence of $^{14}NH_3$ can be related to the contamination or by absorption of atmospheric $N_2$.[28] From the observation, we confirmed that $Mo_2C$ nanosheets protect the host InGaN NWs from the electrolyte and control the $N_2$ leaching on InGaN NWs (figure S2).

Next to the performance, stability is the vital parameter in the search for an ideal photoelectrode for the PEC NRR applications. 6 cycles of CA analysis at the applied bias voltage of -0.2 V vs. RHE were performed (Figure 3 h,i and S14). Finally, post morphological characterization of Mo$_2$C/InGaN NWs confirms the crack-free presence of Mo$_2$C nanosheets over the InGaN NWs and are well in match with the as-prepared samples (Figure S15). Further, it supports the high acid-stable nature of Mo$_2$C to safeguard the host InGaN for long-term NRR applications.

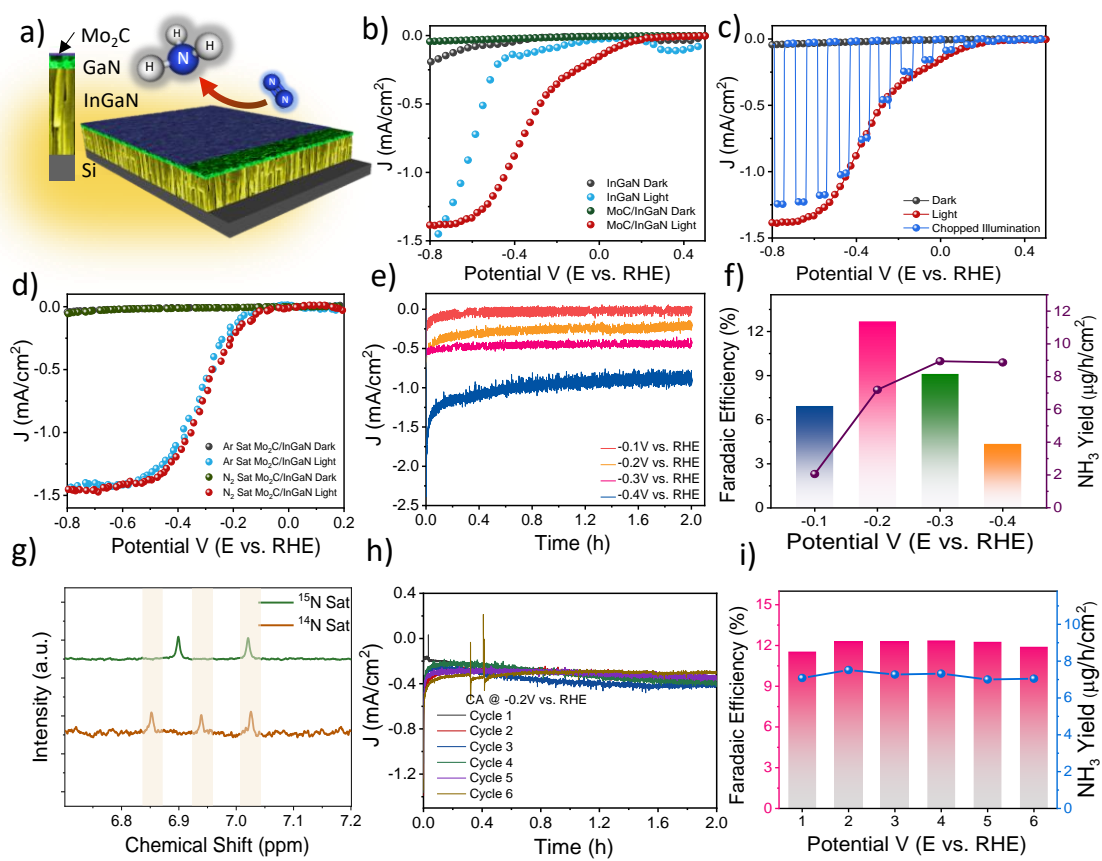

Figure 3: a) Schematic of Mo$_2$C/GaN/InGaN NWs for PEC N$_2$ reduction, b) J-V plot of GaN/InGaN NWs and Mo$_2$C/GaN/InGaN NWs with dark and illumination conditions under ambident condition and 0.05 M H$_2$SO$_4$ as the electrolyte, c) J-V plot of Mo$_2$C/GaN/InGaN NWs under chopped illumination and ambient condition, d) J-V plot of Mo$_2$C/GaN/InGaN NWs under Ar and N$_2$ saturation condition, e) CA analysis of Mo$_2$C/GaN/InGaN NWs at various applied potential vs. RHE under N$_2$ saturated condition, f) Graphical representation of FE and NH$_3$ yield of Mo$_2$C/GaN/InGaN NWs at various applied potential vs. RHE, g) NMR labeling of Mo$_2$C/GaN/InGaN NWs under $^{14}$N$_2$ and $^{15}$N$_2$ feed gas condition, h) stability analysis of Mo$_2$C/GaN/InGaN NWs with 6 cycles of CA at -0.2 V vs. RHE with N$_2$ saturation and i) Graphical representation of 6 cycle stability analysis FE and NH$_3$ yield of Mo$_2$C/GaN/InGaN NWs at an applied bias of -0.2 V vs. RHE.

The effect of incorporating the GaN buffer layer was investigated with interfacial band alignment XPS analysis and provided in figure 4. The point of linear extrapolation of leading-edge in XPS spectra was taken as valence band maxima (VBM). The elemental core level energies were acquired by high-resolution spectra and extracted by Gaussian-Lorentzian fittings (Figure S16). Figure 4a presents the Ga 2p$_{3/2}$ core level and the VBM spectra of p-type GaN was identified to be 1116.41 eV, and 0.68 eV, respectively. Similarly, the Mo 3d$_{5/2}$ and VBM of Mo$_2$C thin film can also be determined from figure 4(b) as 228.47 eV and -0.05 eV, respectively. Figure 4(c) depicts the core level energies of Ga 2p$_{3/2}$ and Mo 3d$_{5/2}$ of Mo$_2$C/p-GaN, which were measured to be 1117.08 eV, and 228.38 eV, respectively. The corresponding valence band offsets ($\Delta E_v$) can be calculated as ~ 1.49 eV using the following equation.[28]

$$\Delta E_v = \left(E_{Mo\ 3d_{5/2}}^{Mo_2C} - E_{VBM}^{Mo_2C}\right) - \left(E_{Ga\ 2p_{3/2}}^{InGaN} - E_{VBM}^{InGaN}\right) + \left(E_{Ga\ 2p_{3/2}}^{Mo_2C/InGaN} - E_{Mo\ 3d_{5/2}}^{Mo_2C/InGaN}\right) \text{---(eq 4)}$$

For a better comparison, band alignment studies were also calculated for Mo$_2$C/InGaN heterostructure and the core level BE spectrum is provided in supporting information Figure S17. The obtained band alignment InGaN/Mo$_2$C and GaN/Mo$_2$C is presented in figure 4d-e, and the corresponding schematic of the band bending of the Mo$_2$C/GaN heterojunction is shown in figure 4f. The bandgap of Mo$_2$C (1.64 eV) is extracted from the Tauc plot fitting of its UV-Vis diffusive reflectance spectrum (figure S18). From the observations, the Mo$_2$C/GaN interface was found to exhibit a typical type-I heterojunction. Interestingly, the small conduction band offset (0.29 eV) and the large VBO (1.49 eV) allow electrons to freely transport from GaN to Mo$_2$C while separating holes at the valence band barrier. From the analysis, we confirmed the incorporation of GaN buffer layer aids the seamless transfer of charge carriers from the InGaN NWs to Mo$_2$C nanosheets.

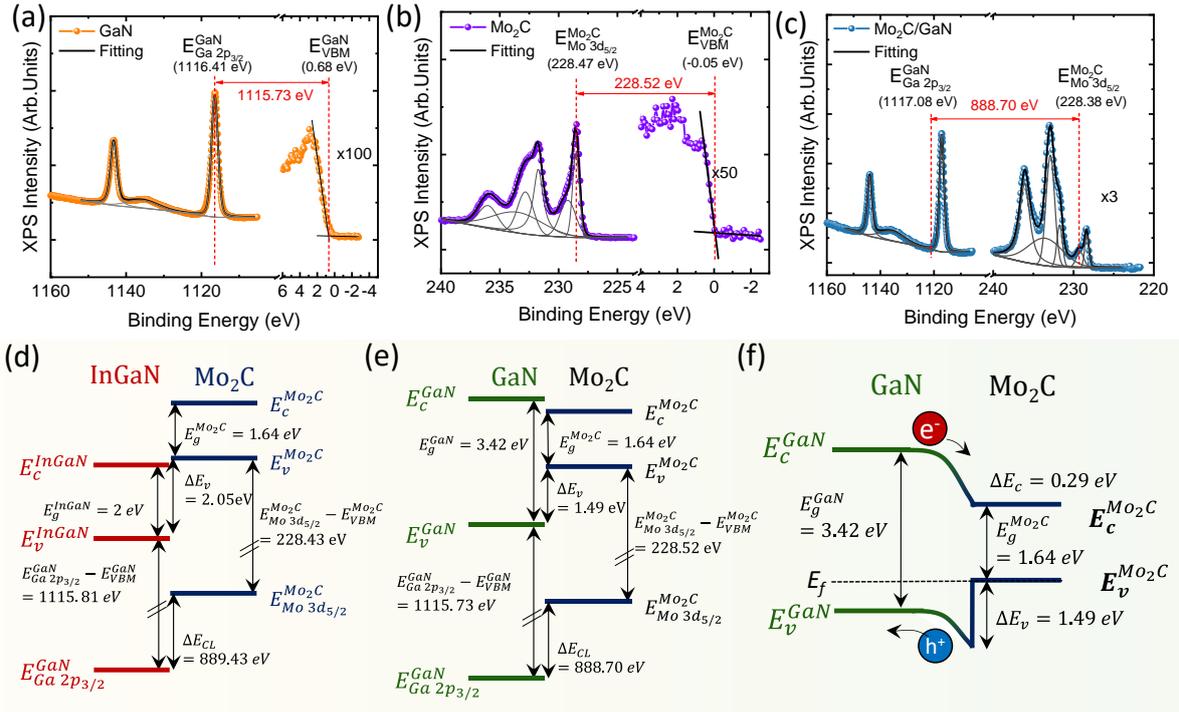

Figure 4: XPS band edge position determination: a) BE spectrum of GaN, b) BE spectrum of Mo₂C, c) BE spectrum of Mo₂C coated GaN, d-e) Energy band diagram of InGaN/Mo₂C and GaN/Mo₂C and f) schematic representation of charge transfer from GaN to Mo₂C.

Overall, the enhanced and stable solar NRR performance of the Mo₂C/InGaN NWs heterostructure can be attributed to

i) Wide range of visible region solar absorption ability of the InGaN NWs

ii) Coalescence top GaN layers host the Mo₂C for uniform coating of ultra-thin Mo₂C layer.

iii) Abundant active catalytic sites and unique *d* band electronic configurations of Mo₂C co-catalyst

iv) GaN buffer layer favors the charge transfer with a suitable band edge position between the InGaN and Mo₂C.

v) Strong acid durability nature of Mo₂C protects the host from harsh electrolyte conditions and avoids N leaching for InGaN NWs.

**Conclusion:**

In summary, Mo₂C/GaN/InGaN NWs heterostructure was fabricated for efficient PEC NRR application. The proposed heterostructure exhibits the maximum FE of 12% and NH₃ yield of 8.9 µg/h/cm$^2$. Besides, the Mo$_2$C/GaN/InGaN heterostructure displays excellent stability up to 6 cycles and high selectivity for ammonia was evidenced by hydrazine estimation. This rich performance of the Mo$_2$C/GaN/InGaN heterostructure is attributed to the high solar absorption cross-section of the InGaN NWs with optimized bandgap, suitable band edge position of GaN buffer layer for efficient charge transport and rich active sites of Mo$_2$C for N$_2$ reduction. In addition to that, the high inert nature of the Mo$_2$C protects the photocathode material from harsh electrolyte conditions for long-term stability.